\documentclass[12pt]{iopart}
\usepackage[T1]{fontenc}
\usepackage{ae,aecompl}
\usepackage{graphicx}
\usepackage{iopams} 
\begin{document}
\title[Persistence of mean-field features in the energy spectrum of small BEC arrays]{Persistence of mean-field features in the energy spectrum of small arrays of Bose-Einstein condensates}

\author{P Buonsante\dag ~\footnote[3]{To whom correspondence should be addressed (pierfrancesco.buonsante@polito.it)}, R Franzosi\ddag\ and V Penna\dag\ }

\address{\dag\ Dipartimento di Fisica and U.d.R. I.N.F.M., Politecnico di Torino, C.so Duca degli Abruzzi 24, I-10129 Torino, Italia}

\address{\ddag\ Dipartimento di Fisica, Universit\`a degli Studi di Pisa, I.N.F.N. sezione di Pisa, and I.N.F.M. U.d.R. di Pisa, Via Buonarroti 2, I-56127 Pisa, Italia}

\begin{abstract}
The Bose-Hubbard Hamiltonian capturing the essential physics of the arrays of interacting Bose-Einstein condensates is addressed, focusing on arrays consisting of two (dimer) and three (trimer) sites. In the former case, some results concerning the persistence of mean-field features in the energy spectrum of the symmetric dimer are extended to the asymmetric version of the system, where the two sites are characterized by different on-site energies. Based on a previous systematic study of the  mean-field limit of the trimer, where the dynamics is exhaustively described in terms of its fixed points for every choice of the significant parameters, an interesting mapping between the dimer and the trimer is emphasized and used as a guide in investigating the persistence of mean-field features in the rather complex energy spectrum of the trimer. These results form the basis for the systematic investigation of the purely quantum trimer extending and completing the existing mean-field analysis. In this respect we recall that, similar to larger arrays, the trimer is characterized by a non-integrable mean-field dynamics featuring chaotic trajectories. Hence, the correspondence between mean-field fixed points and quantum energy levels emphasized in the present work may provide a key to investigate the quantum counterpart of classical instability.



\end{abstract}




\section{Introduction}
\label{S:intro}
In a recent publication \cite{A:PFLet1}, the present authors reported an exhaustive study of the mean-field (MF) dynamics of a system of three interacting Bose-Einstein condensates. Such a system, henceforth  referred to as {\it trimer}, deserves attention since it can be considered a paradigm for larger arrays. Indeed, similar to longer chains and more in general to larger arrays, its MF dynamics is non-integrable and its phase space is characterized by extended regions of instability. On the other hand, the system is still sufficiently small to allow a thorough analytical study, which provides a valuable operational tool in both interpreting experimental observations and identifying significant experimental setups triggering macroscopic  effects \cite{A:PFLet1,A:lungo}. Furthermore, such exhaustive investigation makes the trimer a candidate benchmark for verifying the effectiveness of the MF description, which, being essentially equivalent to the Bogolubov approximation, is expected to provide a reliable picture of the dynamics of an array of interacting BECs for sufficiently large boson populations.

As to the experimental realization of interacting BEC arrays, optical fragmentation allowed to attain 1D \cite{A:BAnderson,A:Pedri,A:Morsch}, 2D \cite{A:Greiner} and even 3D \cite{A:Greiner02} lattices consisting of a rather large number of sites. 
Though Bose-Einstein condensation has been already achieved both in an optomagnetic \cite{A:Andrews} and, more recently, in an entirely magnetic \cite{A:Thomas,A:Tiecke} double-well trap, microtrap technology \cite{A:Reichel,A:Ott} is probably the most promising approach for realizing the small systems addressed by the present article, as well as any complex array characterized by a trapping potential lacking the typical periodicity of optical lattices. 

We will focus on two small systems, the {\it asymmetric dimer}, i.e. a condensate trapped into an asymmetric double well trap, and the {\it open trimer} of reference \cite{A:PFLet1}, where  an energy offset between the central and the lateral wells of the trapping potential is present (a multiwell trapping potential characterized by uneven well depths may possibly be achieved within a supercell of a suitable optical superlattice \cite{A:Roth,A:Sukhorukov}).
Our aim is to compare the spectrum of the quantum Hamiltonians describing these systems with the characteristic energies of the relevant mean-field picture, in the spirit of reference \cite{A:Aubry}. There Aubry and co-workers illustrated a remarkable relation between the energy of the mean-field saddle point (separatrix) of the symmetric dimer  and some features of the spectrum of the quantum counterpart of the same system. 

In the following we will extend these results to the more general case of the asymmetric dimer. Subsequently, exploiting the exact mapping between the mean-field picture of the dimer and that of the integrable sub regime of the trimer, which we show to apply at a qualitative level in the purely quantum treatment of the systems, we will point out some more interesting manifestations of MF features in the spectrum of the trimer.

Before proceeding with our analysis, we briefly recall some generalities. The essential physics of an array of coupled BECs  is captured by the Bose-Hubbard model \cite{A:Jaksch}, describing a gas of interacting bosons hopping across the sites of an ambient lattice. The relevant Hamiltonian reads $H=\sum_j [U n^2_j -v_j n_j -T/2 \sum_{k\sim j}(a_j^+ a_k+a_k a^+_j)]$, where $a^+_j$ ($a_j$) is the creation (annihilation) operator on site $j$, $n_j =a_j^+ a_j$ counts the number of bosons at the same on site, $U$ is the on-site interaction parameter, $v_j$ is a the energy offset of site $j$,  $T$ is the hopping parameter and the subscript $k\sim j$ restricts the sum to the nearest neighbours of site $j$.

A mean-field treatment of the Bose-Hubbard model can be achieved by means of a time dependent variational principle based on Glauber's coherent states \cite{A:Amico}. According to such picture each site in the array is described by a macroscopic complex variable $z_j=\sqrt{\rho_j}\, \rme^{\rmi \phi_j}$ whose square modulus $\rho_j$ and argument $\phi_j$ represent the population and phase associated to the condensate sitting at that site, respectively. 
Barring constant terms, the effective MF Hamiltonian has the form ${\cal H}=\sum_j [U |z_j|^4 -v_j |z_j|^2 -T/2 \sum_{k\sim j}(z_j^* z_k+z_j z_k^*)]$ and, equipped with the Poisson brackets among the (canonical) macroscopic variables,  $\{z_j,z^*_k\}=\delta_{jk}$ determines the dynamics of the system. \footnote{The set of equations governing the dynamics of the system are basically the same  discrete nonlinear Sch\"rodinger equations (DNLEs) deriving from the  application of the tight-binding approximation to the Gross-Pitaevski equation describing the evolution of a fragmented BEC \cite{A:TrombL}. Reference \cite{A:Andersen93} reports on some analytical solutions of similar DNLSEs on small arrays such as the ones presently under concern.}. The essential features of the dynamics are determined by the phase space structure, which can be analyzed working out the location and the stability character of the corresponding fixed points. Notice that since $H$ features only two-boson operators, it commutes with the the total number operator $n = \sum_j n_j$. Likewise, the total population $\rho = \sum_j \rho_j$ is a constant of the motion mirroring the global phase invariance of $\cal H$. Hence the fixed point equations have the form $\{z_j,{\cal H}-\mu\,\rho\}=0$, where $\mu$ is a Lagrange multiplier constraining the total number of bosons. Though the solutions of the fixed point equations are in principle complex vectors, they can be safely restricted to real values, with the only exception of the topological solutions  \cite{A:Casetti,A:Paraoanu} which are to be found in cyclically invariant structures (i.e. homogeneous closed chains and their higher dimensional generalizations). Since the structures we will consider are neither cyclic nor homogeneous, in solving the fixed point equations we will set $z_j = \sqrt{\rho}\, x_j$, where $x_j \in \mathbb{R}$ and $\sum_j x_j^2=1$.


 \section{The asymmetric dimer}
\label{S:dim}
The Bose-Hubbard Hamiltonian for an asymmetric dimer has the form
 \begin{equation}
 \label{E:H2}
 H_{\rm d} =  U_{\rm d}\left(n^2_\ell+n^2_r\right) -\frac{w_{\rm d}}{2} (n_\ell - n_r) -\frac{T_{\rm d}}{2}\left(a_\ell\,a_r^++a_\ell^+\,a_r\right)
 \end{equation}
where the subscripts $\ell$ and $r$ denote the left and the right sites of the structure, respectively, and $w_{\rm d}$ is a local energy offset deriving from the asymmetry of the double-well trap confining the BEC. 
As we briefly recalled above, the MF counterpart of Hamiltonian \eref{E:H2} is $ {\cal H}_{\rm d} =  U_{\rm d}(|z_\ell|^4+|z_r|^4) -\frac{1}{2}w_{\rm d} (|z_\ell|^2 - |z_r|^2) -\frac{1}{2} T_{\rm d}(z_\ell\,z_r^*+z_\ell^*\,z_r)$, and determines the  evolution of the macroscopic semiclassical variables through the (integrable) equations of motion
\begin{equation}
\label{E:ddyn}
\fl \rmi \hbar \dot z_\ell = \left(2 U_{\rm d} |z_\ell|^2 -\frac{w_{\rm d}}{2}\right) z_\ell -\frac{T_{\rm d}}{2} z_r, \qquad \rmi \hbar \dot z_r = \left(2 U_{\rm d} |z_r|^2 +\frac{w_{\rm d}}{2}\right) z_r -\frac{T_{\rm d}}{2} z_\ell \\
\end{equation} 
According to the discussion in the previous section, and after introducing the conveniently rescaled parameters
\begin{equation}
\nu_{\rm d} = \frac{w_{\rm d}}{U_{\rm d} \rho_{\rm d}}, \quad \tau_{\rm d} = \frac{T_{\rm d}}{U_{\rm d} \rho_{\rm d}}, \quad \mu_{\rm d} = \frac{\chi_{\rm d}}{U_{\rm d} \rho_{\rm d}}
\end{equation}
where $\rho_{\rm d} = |z_\ell|^2+|z_r|^2= \rho_{\rm d} (x_\ell^2+x_r^2)$, the consequent fixed point equations read
\begin{equation}
\label{E:dfp}
 \left(2 x_\ell^2 -\frac{\nu_{\rm d}}{2}-\mu_{\rm d}\right) x_\ell -\frac{\tau_{\rm d}}{2} x_r =0, \qquad  \left(2 x_r^2 +\frac{\nu_{\rm d}}{2}-\mu_{\rm d}\right) x_r -\frac{\tau_{\rm d}}{2} x_\ell = 0
\end{equation}
Notice that $\tau_{\rm d}$ and $\nu_{\rm d}$  are the only  significant parameters, whereas $\mu_{\rm d}$ has an auxiliary function and its value is different for different fixed points. 
In the following we will consider $|\nu_{\rm d}|$'s up to the order of $2$, corresponding to relative energy offsets of the order of $2\, U\, \rho_{\rm d}$. If large total populations are involved, the limit situation may correspond to an overly asymmetric double well potential. Notice however that comparable relative energies are easily attained in the MF integrable subregime of the trimer, which can be described in terms of an asymmetric dimer, as we will discuss in the next section.

Depending on whether $|\nu_{\rm d}|$ is  smaller or larger  than $(2^{2/3}-\tau_{\rm d}^{2/3})^{3/2}$, \eref{E:dfp} have either four or two solutions. In the former case there are two maxima, a saddle point and a minimum. As $|\nu_{\rm d}|$ grows, the local maximum and the saddle get closer and closer, until they collapse and disappear when $|\nu_{\rm d}|$ crosses the ($\tau_{\rm d}$-dependent) critical value \footnote{Notice that the coalescence of the (possibly local) maximum and the saddle can be likewise achieved also through a variation of $\tau_{\rm d}$, and occurs at a $|\nu_{\rm d}|$-dependent critical value \cite{A:Aubry,A:sp_pr}. A similar coupling-driven change in the semiclassical phase space of a 2D BEC array has been recently proposed as a mechanism to trigger an abrupt delocalization in the system \cite{A:Kalosakas02}.}. We mention that the presence of two maxima is an aspect of the mirror-symmetry breaking induced by nonlinearity, leading to  interesting phenomena such as self-trapping and $\pi$-phase oscillations, that are investigated in reference \cite{A:Raghavan} for the symmetric dimer. 

Aubry and co-workers \cite{A:Aubry} focused on the symmetric dimer ($w_{\rm d}=0\;\Rightarrow\;\nu_{\rm d}=0$), and found that when the MF saddle point (separatrix) is present, i.e. for $\tau_{\rm d}<2$, its energy is strictly related to some interesting features of the spectrum of $H_{\rm d}$. They noticed that the density of states of  $H_{\rm d}$ features a sharp peak centered around the saddle point energy. Furthermore this MF energy marks the boundary between two regions of the spectrum characterized by significantly different magnitudes of the splitting energy of the doublets arising from the mirror symmetry of the symmetric dimer. More precisely the energy splitting experiences a drop of several orders of magnitude as the average energy of the doublet becomes larger than the threshold value marked by the saddle point energy. 
Further analytical considerations about the splitting pattern of the quantum dimer doublets and its relation with the saddle point energy of the corresponding mean-field model can be found in reference \cite{A:sp_pr}.
Notice that the magnitude of the splitting between even-odd doublets of the symmetric dimer is also related to the characteristic time-scale for the symmetry-restoring quantum tunneling between semiclassically self-trapped states \cite{A:Raghavan2,A:Smerzi,A:sp_pr}.

Here we show that the results by Aubry and co-workers can be extended and generalized to the asymmetric dimer, where  $w_{\rm d}$ (and of course $\nu_{\rm d}$) has a finite value. To this end it proves useful to plot the energies of the MF fixed points of ${\cal H}_{\rm d}$ versus the energy offset among the two sites of the dimer, $w_{\rm d}$, which gives a measure of its asymmetry.
This can be accomplished recasting the dynamics of the trimer in terms of the alternative set of canonical variables (See e.g. reference \cite{A:IJMP} for further details about this alternative canonical representation).  $\rho_{\rm d}=|z_1|^2+|z_2|^2$, $D=|z_1|^2-|z_2|^2$, $\theta=(\phi_1-\phi_2)/2$, $\psi=(\phi_1+\phi_2)/2$, whose significant Poisson brackets are $\{D,\theta\}=\{\rho_d,\psi\}=\hbar^{-1}$. This way, the relevant dynamic equations can be used to give a parametric representation allowing to plot the fixed point energies versus $w_{\rm d}$ for a given value of the hopping amplitude. Since it is interesting to compare situations characterized by different total populations, we provide this parametric representation in terms of the rescaled parameters $\tau_{\rm d}$, $\nu_{\rm d}$,  
\begin{equation}
\label{E:Pne}
\fl \nu_{\rm d}(d;\tau_{\rm d})= -2\, d\left(1+\frac{\tau_{\rm d}\,\cos{2 \theta}}{2\sqrt{1-d^2}} \right),\qquad E_{\rm d}(d;\tau_{\rm d})= \frac{(1-d^2)^\frac{3}{2}-\tau_{\rm d} \cos 2 \theta}{2\sqrt{1-d^2}},
\end{equation}
where $E_{\rm d}$ is the energy in units of $U \rho_{\rm d}^2$,  $d=D/\rho_{\rm d} \in [-1, 1]$ is the rescaled population imbalance \cite{A:Raghavan},
 and $\theta$ is either zero or $\pi/2$. The former choice pertains to the minimum fixed point no matter for the value of the hopping amplitude. As we discussed above, when $\tau_{\rm d}> 2$ there is only one more point, a maximum, independent of the value of the energy offset $|\nu_{\rm d}|$. In this case \eref{E:Pne} the remaining choice for the angular variable, $\theta=\pi/2$, describes the energy of the maximum on the entire domain, $d\in [-1,1]$. If, conversely,  $\tau_{\rm d}< 2$, the choice  $\theta=\pi/2$ pertains to different fixed points depending on the domain of evaluation. More in detail, the energy of the saddle, local maximum and absolute maximum are obtained evaluating  \eref{E:Pne} with $|d|$ belonging to the interval $[0,d_1]$, $[d_1,d_2]$ and $[d_2,1]$, respectively, where $d_1=\sqrt{1-(\tau_{\rm d}/2)^{2/3}}$ and  $d_2=\sqrt{1-(\tau_{\rm d}/2)^2}$ .

 \begin{figure}
 \begin{center}
 \begin{tabular}{cc}
 \includegraphics[width=7.5cm]{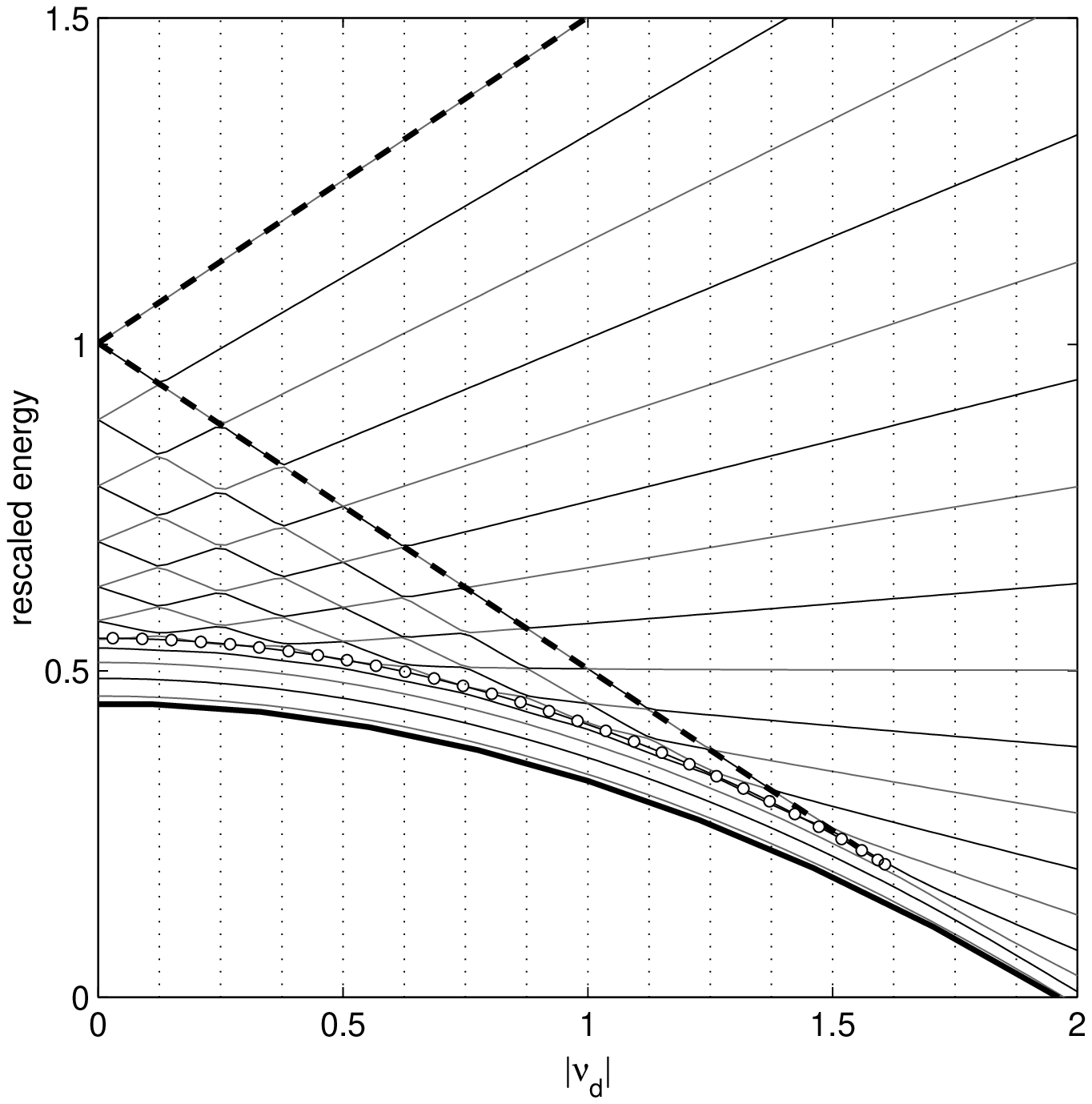}&
 \includegraphics[width=7.5cm]{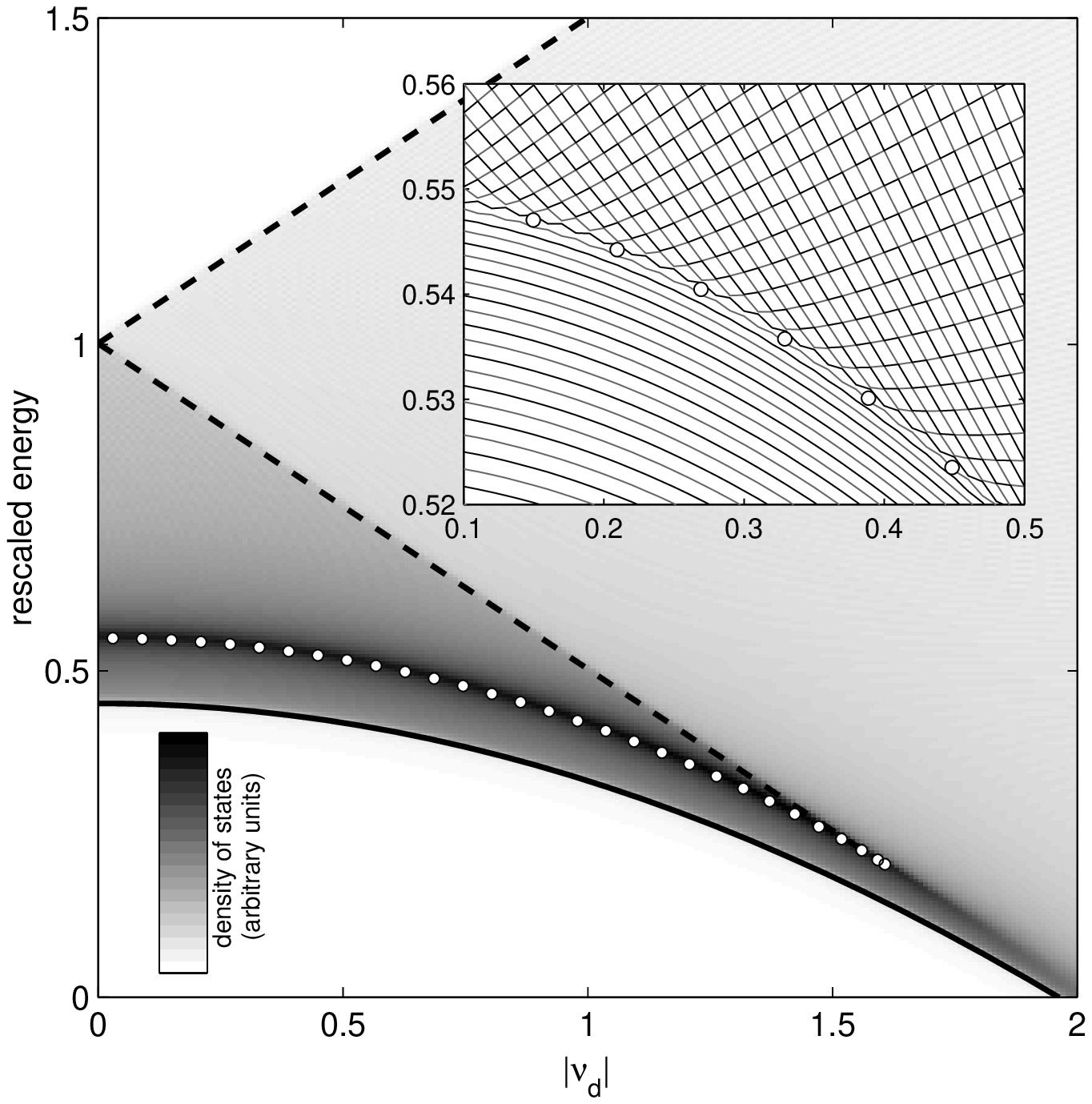}
 \end{tabular}
 \caption{\label{F:dim} Comparison between the mean-field and quantum characteristic energies for the asymmetric dimer for $\tau_{\rm d}=0.1$. {\bf Left panel}. Thin solid lines: energy levels for a dimer containing $\rho_{\rm d}=16$ bosons. Alternating colours were used to emphasize the avoided level crossing. Thick black line: MF minimum; Thick black dashed lines: MF maxima. Circles: MF saddle. {\bf Right panel}. The MF energies are compared to the density of the states for a dimer containing $\rho_{\rm d}=200$ bosons. The inset shows the spectrum and the energy of the MF saddle in a small region of the figure plane.}
 \end{center}
 \end{figure}
The  circles, the thick solid line and the thick dashed lines appearing in both panels of figure \ref{F:dim} were obtained as described above, and show the behaviour of the energy of the MF  saddle, minimum and maxima, respectively, for  $\tau_{\rm d}=0.1$. In  the left panel of the same figure we also plotted the energy levels of $H_{\rm d}$ for a total population of  $\rho_{\rm d}=16$ bosons (thin solid lines), using alternating colors for emphasizing the avoided level crossing feature. Notice indeed that at a discrete set of values of the energy offset ($w_{\rm d}= 2 k U_{\rm d} $, $k=1,\ldots, \rho_{\rm d}$), signalled by the vertical dotted lines, some pairs of adjacent energy levels become so close that, if plotted with the same color, they would give the illusion of crossing as they actually do when $\tau_{\rm d}=0$. If conversely  $\tau_{\rm d}>0$, the adjacent energy levels never cross\footnote{We mention that the avoided level crossing is discussed to some extent in reference \cite{A:Flach}. However there the varying parameter is the hopping amplitude between the sites of a dimer and a third site lacking of the repulsive term.}, but rather they ``bounce'' on each other, so that their behaviour with varying $\nu_{\rm d}$ resembles the trajectory of particles repelling each other (see e.g. \cite{A:Nakamura}). Hence, whenever the energy offset between the wells is an even multiple of $U_{\rm d}$ ($\nu_{\rm d}= 2 k/\rho_{\rm d}$), the situation becomes similar to that of the symmetric dimer: the couples of degenerate levels possibly present for $\tau_{\rm d}=0$ split for finite values of the hopping parameter. Once again, the extent of the splitting of a given doublet can vary of order of magnitudes depending on whether its average energy is smaller or larger than the MF saddle point energy.
Hence also in the asymmetric case such energy marks a clear boundary between two regions of the spectrum with quite different features. This is rather evident in the inset of the right panel of figure \ref{F:dim}, where we plotted the energy levels for a quantum dimer with a total population of  $\rho_{\rm d}=200$ bosons (black and gray solid lines) in a small region around the saddle point energy (white circles). The main plot of the right panel of figure \ref{F:dim} shows the (Lorentzian broadened) density of states of Hamiltonian \eref{E:H2} for the same total population. It appears evident that also in the asymmetric case the sharp peak of the density of states (dark shading) follows very closely the saddle point energy (white filled circles). These features are observed whenever the hopping amplitude is sufficiently small ($\tau_{\rm d}<2, \, T_{\rm d}<2 U_{\rm d}\, \rho_{\rm d}$) to allow the appearance of a saddle point for some values of $\nu_{\rm d}$.
\section{The trimer}
\label{S:trm}
In reference \cite{A:PFLet1} the present authors analyzed the mean-field dynamics of a trimer described by the following Bose-Hubbard Hamiltonian
\begin{equation}
\label{E:H3}
 H_{\rm t}= U_{\rm t}\sum_{j=1}^3 n_j^2 + \frac{w_{\rm t}}{2}\left(n_1- n_2+n_3\right)
-\frac{T_{\rm t}}{2}[(a_1+a_3)\,a_2^++(a_1^++a_3^+)\,a_2]
\end{equation}
where both the relative offset of the central well, $w_{\rm t}$, and the hopping amplitude $T_{\rm t}$ were regarded as adjustable parameters. The results of this study were gathered into two {\it stability diagrams}  summarizing, for any choice of the significant parameters, the character and linear stability of the entire set of MF fixed points, and hence describing the essential features of the phase space structure. This knowledge would allow to produce the trimeric counterparts of figure \ref{F:dim} with no further effort other than the diagonalization of Hamiltonian \eref{E:H3}. Notice however that  the size of this operator scales as $(\rho_{\rm t}+1)(\rho_{\rm t}+2)/2$, which could be so large to make analytical diagonalization impracticable even for a relatively small total boson population $\rho_{\rm t}$. Thus the only readily available analytical information to provide a systematic description of the quite intricate spectrum of the trimer (see e.g. figure \ref{F:Qmap}) is the $Z_2$ symmetry of \eref{E:H3}, allowing to classify its energy levels according to their parity.
 
This is why here we prefer to proceed discussing a close relation between the fixed point of the trimer and those of the dimer which will turn out to persist also at the quantum level, allowing us to use the results reported in section \ref{S:dim} as a guide to gain deeper insight in the spectrum of the trimer. 
\subsection{The mean-field picture of the trimer and its analogy to the symmetric dimer}
\label{S:MFmap}
The effective Hamiltonian ensuing from \eref{E:H3},  $ {\cal H}_{\rm t} =  U_{\rm t} \sum_j |z_j|^4 +\frac{1}{2}w_{\rm t} (|z_1|^2 - |z_2|^2+|z_3|^2) -\frac{1}{2} T_{\rm t}[z_2^*(z_1+z_3)+z_2 (z_1^*+z_3^*)]$, yields the following dynamic equations
\begin{equation}
\label{E:tdyn}
\fl \rmi \hbar \dot z_j = \left(2U |z_j|^2+\frac{w_{\rm t}}{2}\right)z_j-\frac{T_{\rm t}}{2}z_2,\qquad \rmi \hbar \dot z_2 = \left(2U |z_2|^2-\frac{w_{\rm t}}{2}\right)z_2-\frac{T_{\rm t}}{2} \left(z_1+z_3\right)
\end{equation}
where the first equation applies to both of the lateral sites, $j=1,3$. This feature is a consequence of the mirror symmetry that ${\cal H}_{\rm t}$ inherited from  $H_{\rm t}$, and allows room for an integrable sub regime in the generally non-integrable dynamics of the trimer \cite{A:PFLet1,A:lungo,A:Lphys3}. Notice indeed that the initial condition $z_1=z_3$ is propagated in time by \eref{E:tdyn}. As we will discuss in some detail in the following, this sub regime of the trimer can be mapped exactly onto the asymmetric dimer considered in the previous section, which is why it is also referred to as {\it dimeric sub regime} of the trimer. However, since the analogy with the dimer applies more in general to the energies of most of the fixed points of the trimer, either exactly or approximately, we will refrain from assigning the adjective ``dimeric'' exclusively to the integrable sub regime of the trimer and the relevant fixed points. To avoid confusion, the latter will be referred to as {\it symmetric fixed points} (SFPs) alluding to their feature $z_1=z_3$. Likewise we will term {\it asymmetric fixed point} (AFP) any fixed point such that $z_1\neq z_3$, with the only exception of the so-called {\it central depleted well} \cite{A:PFLet1} ($z_1=-z_3$, $z_2=0$) which, having no counterpart in the asymmetric dimer, will not be considered in the following discussion.

The close relation between the integrable sub regime of the trimer and the asymmetric dimer can be emphasized by recasting equations \eref{E:tdyn}, subject to the (dynamically propagated) constraint $z_1=z_3$, in terms of the alternative variables \footnote{Likewise, if the identification $(z_\ell,\,z_r)=(z_1/\sqrt 2,\,z_2)$ is made, the resulting equations have the same form as those describing a dimer with site-dependent on-site interaction \cite{A:Raghavan,A:Lphys3}.}
$d=(|z_1|^2/2- |z_2|^2)/\rho_{\rm t}$ and $\phi=(\phi_1-\phi_2)/2$.
The resulting equations are exactly in the same form as those ensuing  describing the dynamics of an asymmetric dimer with the following parameters
\begin{equation}
\label{E:parmapE}
 U_{\rm d} = \frac{3}{4}U_{\rm t},\qquad  w_{\rm d} = w_{\rm t}-\frac{1}{2}\rho_{\rm t} U_{\rm t},\qquad  T_{\rm d}= \sqrt 2\, T_{\rm t},\qquad  \rho_{\rm d}= \rho_{\rm t}
\end{equation}
Hence, as we mentioned in section \ref{S:dim},  the integrable sub regime of a trimer with no energy offset ($w_{\rm t}=0$) could provide a realization of a rather asymmetric dimer, the corresponding relative offset being proportional to the (possibly very large) total population, $w_{\rm d}=-\frac{1}{2} \rho_{\rm t}  U_{\rm t}$.

\begin{figure}
 \begin{center}
 \begin{tabular}{cc}
 \includegraphics[width=7.5cm]{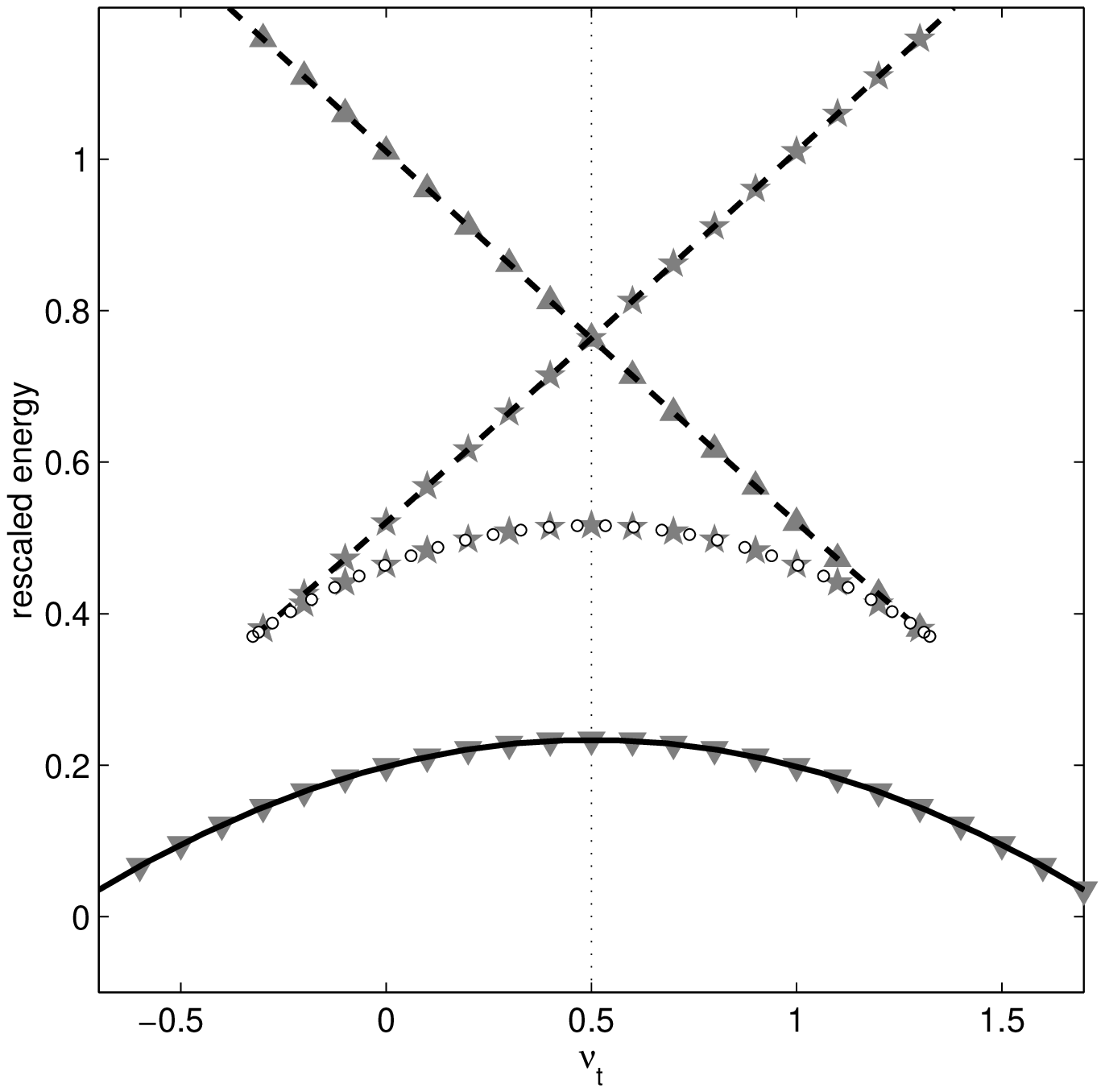}&
 \includegraphics[width=7.5cm]{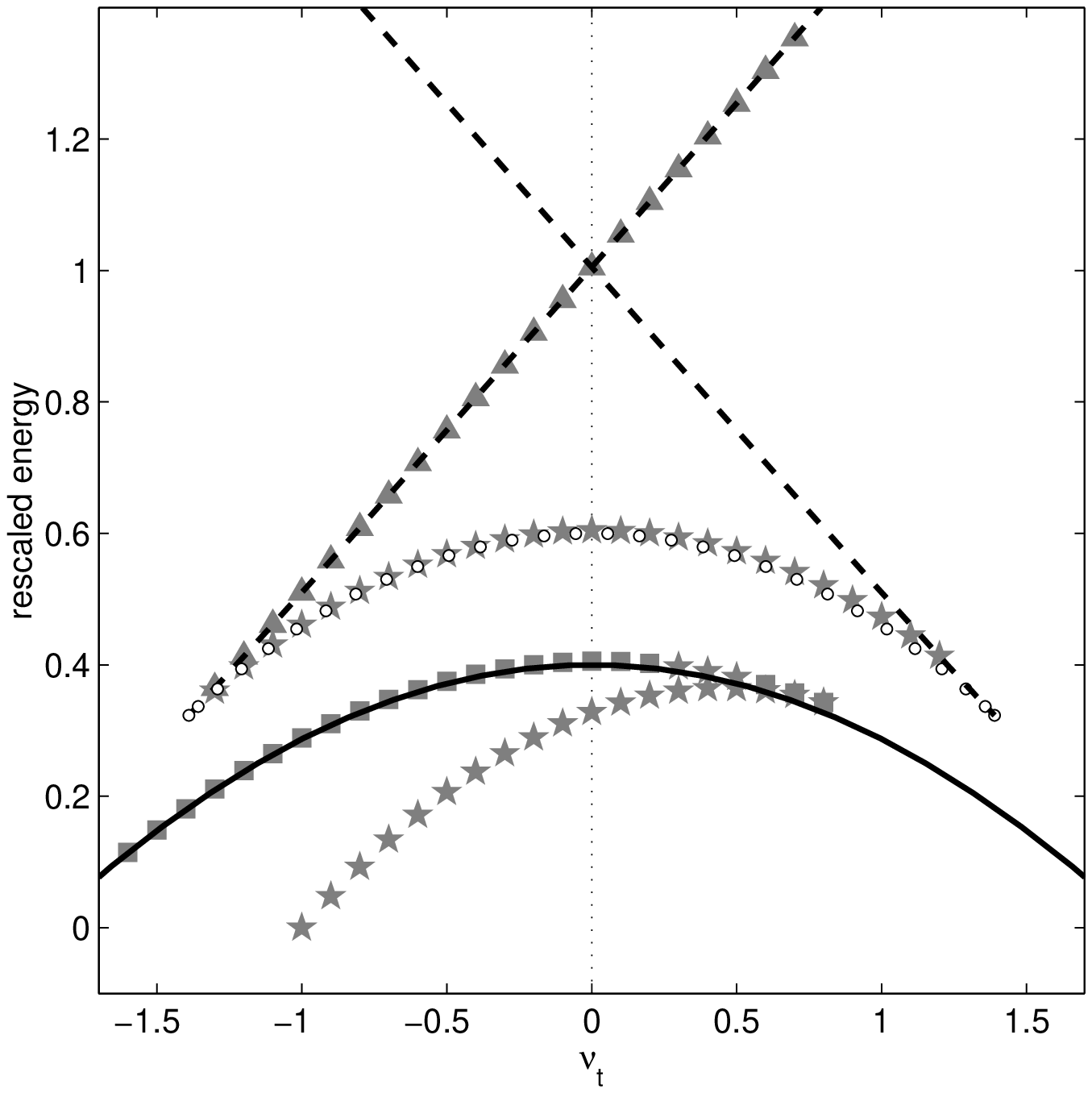}
 \end{tabular}
 \caption{\label{F:MFmap} Trimer-dimer analogy for $\tau_{\rm t}=0.2$. {\bf Left panel}: exact mapping \eref{E:parmapE} between the fixed points of the asymmetric dimer and the SFPs of the trimer. The gray symbols denote the local and stability character of the fixed point as evaluated in the entire phase space. More precisely gray stars, upward and downward triangles denote saddles, (either local or absolute) maxima and minima, respectively. White filled circles, dashed and solid thick black lines denote the same local and stability characters as evaluated within the sub manifold of the phase space pertaining to the integrable sub regime ($z_1=z_3$). {\bf Right panel}: approximate mapping \eref{E:parmapA} between the fixed point of the dimer and the AFP of the trimer. The gray squares denote the stable saddles typical of the non-integrable regime of the trimer \cite{A:PFLet1}, whereas the remaining symbols are the same as in the left panel. The dotted black lines appearing in both panels signal the situation corresponding to a symmetric dimer.}
 \end{center}
 \end{figure}

Here we will not give further analytic details about the study of the trimer dynamics (solution of the ``asymmetric'' fixed point equations, stability analysis), which can be found elsewhere \cite{A:PFLet1,A:lungo}, our present concern being the analogy between the trimer and the dimer.
In this respect we remark that the above described exact mapping applies also to the stability character of the SFP only within the ``symmetric'' phase subspace $z_1=z_3$ pertaining to the integrable sub regime. Indeed, when considering the entire phase space, the SFP corresponding to one of the (stable) maxima of the asymmetric dimer may turn out to be unstable. This is clearly shown in the left panel of figure \ref{F:MFmap}, where we plotted the energies the SFPs, as obtained with the straightforward modifications to equation \eref{E:Pne} ensuing from the mapping \eref{E:parmapE}: $\nu_{\rm t}(d;\tau_{\rm t})=\frac{1}{2}+\frac{3}{4} \nu_{\rm d}(d;\tau_{\rm d})$, $E_{\rm t}(d;\tau_{\rm t})=\frac{3}{4} E_{\rm d}(d;\tau_{\rm d})$, $\tau_{\rm d}=\frac{3}{4\sqrt 2}\tau_{\rm t}$.

As a further, somewhat unexpected, aspect of the dimer-trimer analogy,  the parametric representation \eref{E:Pne} allows to describe also the energies of most of the asymmetric fixed points of the trimer, though approximately and for not exceedingly large values of $\tau_{\rm t}$. An example of this situation is portrayed  in the right panel of figure \ref{F:MFmap}, where we plotted the energy of the AFPs (gray markers) along with those of the fixed points of an asymmetric dimer (black lines, circles) with exactly the same parameters as the trimer: 
\begin{equation}
\label{E:parmapA}
 U_{\rm d} = U_{\rm t},\qquad  w_{\rm d} = w_{\rm t},\qquad  T_{\rm d}= T_{\rm t},\qquad  \rho_{\rm d}= \rho_{\rm t}
\end{equation}
As we mentioned above, the larger the rescaled hopping amplitude $\tau_{\rm t}$ the less satisfactory the overlap between the fixed points of dimer and the AFPs of the trimer. However, the comparison among situations relevant to different hopping amplitudes seems to suggest that a more complex parameter mapping than \eref{E:parmapA} could produce better overlaps. Due to the limited space, here we do not develop this issue, but we rather limit our qualitative discussion to sufficiently small values of the hopping amplitude (up to $\tau_{\rm d}\sim 0.5$) for the simple mapping in \eref{E:parmapA} to give satisfactory results.  In this respect we mention that a closer analysis of the relevant fixed point equations \cite{A:lungo} shows that the dimer analogy is more effective for the AFPs characterized by a very low population of one of the lateral sites. 
A further feature worth noticing is the correspondence between the dimeric minimum (thick solid line) and the stable saddle points \footnote{Recall that, according to the (linear) stability theory, the maxima and the minima are always stable, whereas the saddle point can be either stable or unstable. In the case of the trimer, the analytically predicted stability of the saddle points has been widely confirmed by numerical simulations \cite{A:lungo}.} typical of the non-integrable regime of the trimer (gray squares).  
We conclude this somewhat qualitative analysis of the dimer-trimer correspondence by emphasizing that one of the dimeric maxima (black dashed lines) has no counterpart among the AFPs. As we will briefly discuss in the following, a similar feature appears in the in the purely quantum treatment of the problem, in terms of parity of the energy levels.
\subsection{Quantum persistence of the dimer-trimer analogy}
\label{S:Qmap}
\begin{figure}
\begin{center}
\begin{tabular}{cc}
\includegraphics[width=7.5cm]{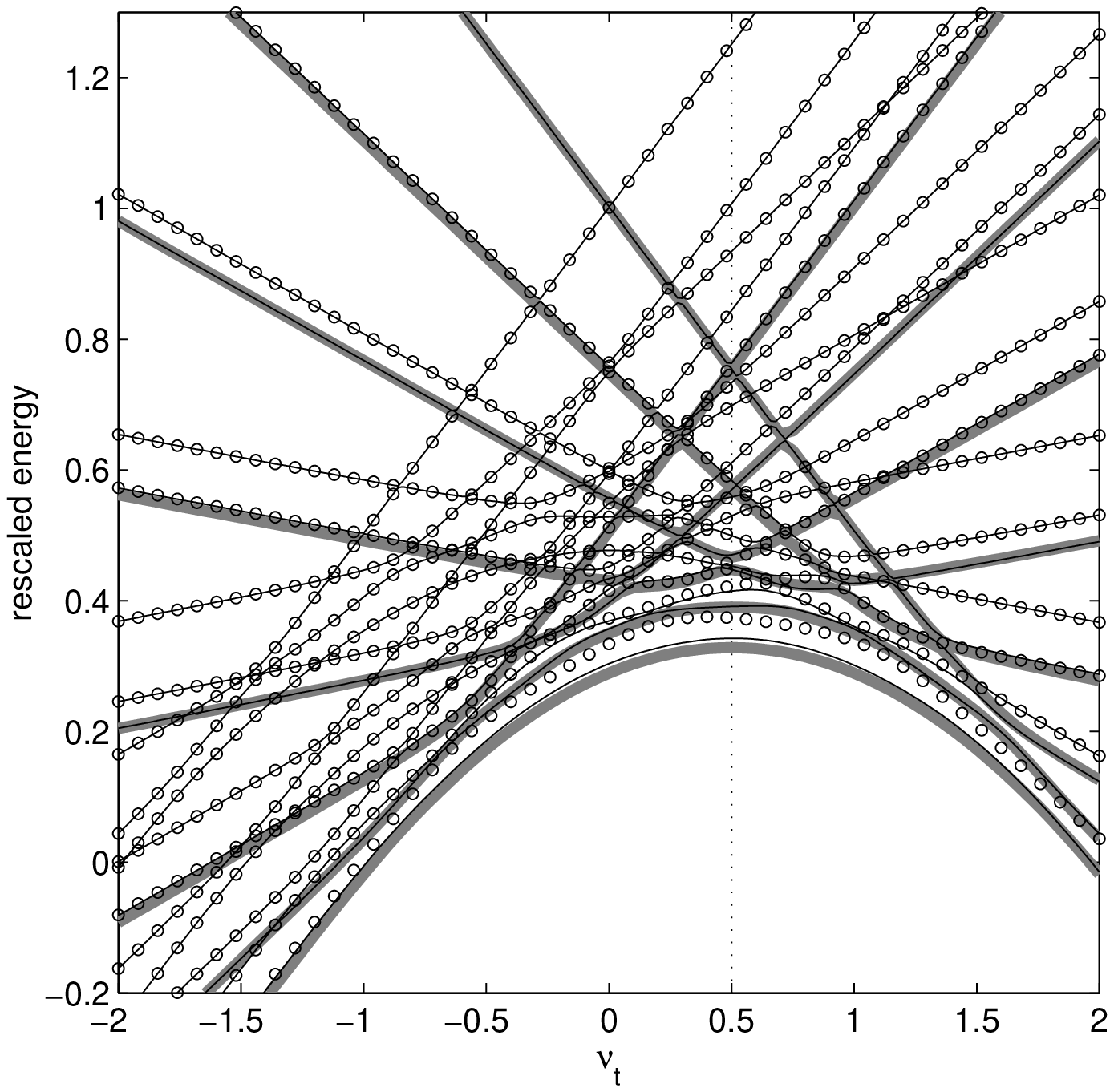}&
\includegraphics[width=7.5cm]{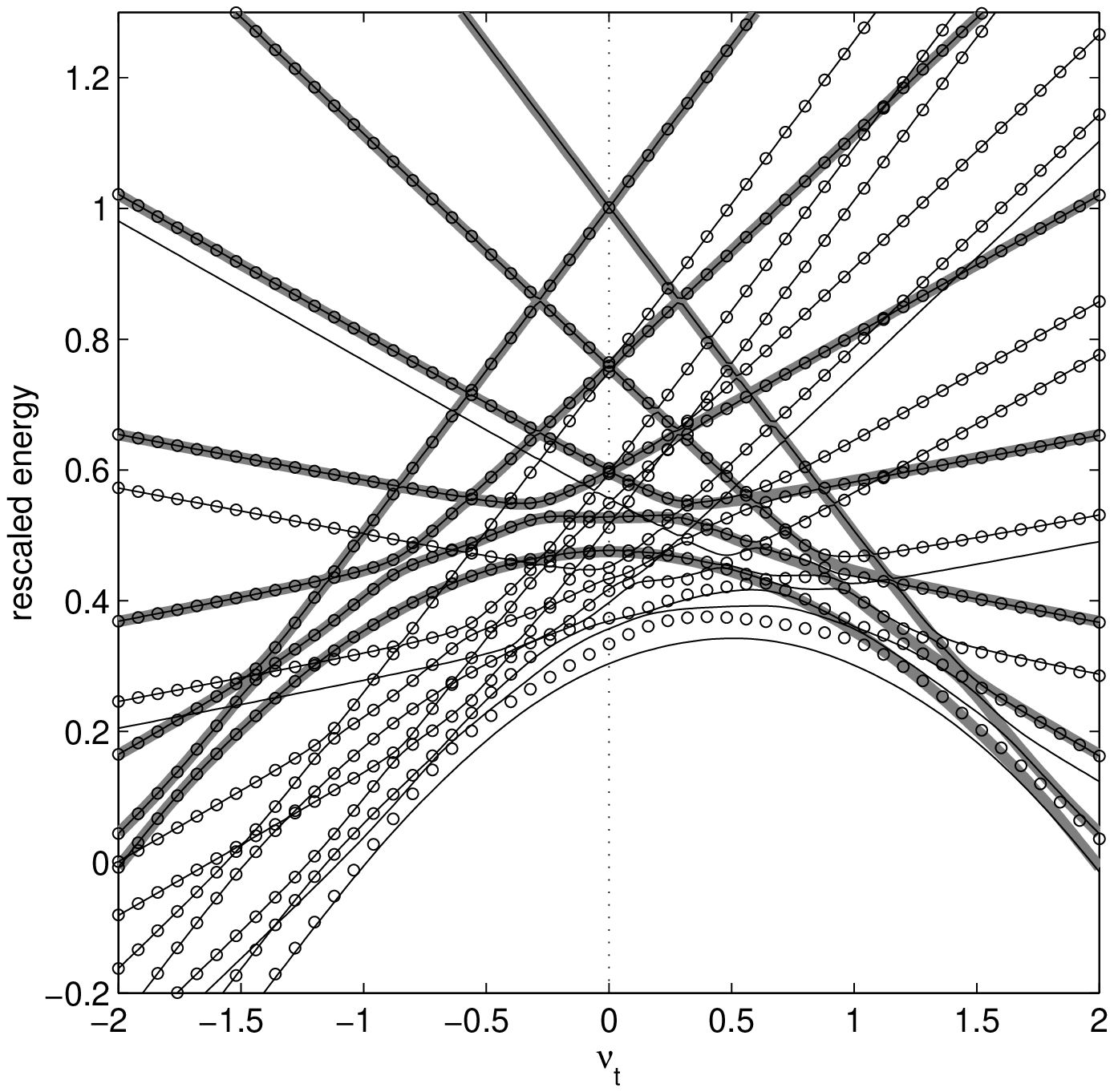}
\end{tabular}
\caption{\label{F:Qmap} Quantum counterpart of figure \ref{F:MFmap} for $\tau_{\rm t}=0.1$ and $\rho_{\rm t}=7$. The thin solid lines and the circles denote even and odd energy levels of the trimer, respectively. The thick gray lines are the levels of the dimeric Hamiltonian \eref{E:H2}. Those in the left panel have been obtained with the parameter choice \eref{E:parmapE} providing the exact mapping between the dynamics of ${\cal H}_{\rm d}$ and the integrable sub regime of ${\cal H}_{\rm t}$. The dimeric levels in the right panel have been obtained according to the approximate mapping \eref{E:parmapA}.}
\end{center}
\end{figure}

As we discussed at the beginning of the present section, the Hilbert space for a trimer is so large to require a numerical diagonalization of the (number conserving) Hamiltonian \eref{E:H3} even for relatively small boson populations $\rho_{\rm t}$. Likewise, the pattern produced by the energy levels of the trimer as the relative energy offset $w_{\rm t}$ is varied is much more complex than its dimeric counterpart (see e.g. figure \ref{F:Qmap}). Some help in the classification of these energy levels comes from the observation that parity is a good quantum number, which casts some light on  the rather complex avoided crossing pattern occurring as $w_{\rm t}$ varies. The crossing of two levels is either forbidden or allowed depending on whether they have the same or different parity. Indeed levels with different parity belong to independent sectors of the Hilbert space, and there is no hindrance to their crossing.

The discussion presented in section \ref{S:MFmap} provides the very interesting
suggestion of interpreting the energy level pattern of the trimer in terms of the spectrum of a dimer with the same population and suitably chosen parameters. Actually, the dimer-trimer analogy persists also at the quantum level, as it is clearly shown in figure \ref{F:Qmap}, where we plotted the 36 levels of a trimer containing $\rho_{\rm t}=7$ bosons along with the spectrum of two different dimers with the same population. The dimeric spectra appearing in the left and right panels of the figure were obtained from the diagonalization of Hamiltonian \eref{E:H2} with the parameter choices in \eref{E:parmapE} and \eref{E:parmapA}, respectively. Notice that in both cases each dimeric level follows quite closely a level of the trimer. In this respect we emphasize that the recognition of dimeric structures in the spectrum of \eref{E:H3} is quite difficult even after the introduction of an alternative canonical scheme incorporating the parity symmetry of the trimer \cite{A:Lphys3}, and involves non-trivial manipulations hardly conceivable except as an afterthought following the MF analysis.

Notice that any definite-parity quantum state yields $\langle n_1-n_3\rangle=\langle a_1^+ a_3-a_3^+ a_1\rangle=0$, which is a possible quantum counterpart of the MF integrability condition $z_1=z_3$\footnote{$\langle a_{\rm e}^+ a_{\rm e}\rangle=0$, where $a_{\rm e} =(a_1+a_3)/\sqrt 2$ makes probably a better candidate. Due to limited space, we refer to \cite{A:Lphys3} for a discussion about this issue.}. This is basically the reason why each dimeric level in the right panel of figure \ref{F:Qmap} overlaps with an odd-even doublet of the trimer: the latter is the minimal set of eigenstates producing a state lacking a definite parity, which is the quantum counterpart of a MF asymmetric configuration. Recall indeed that the dimer under concern is the quantum counterpart of that qualitatively describing the asymmetric fixed points of the trimer.
Conversely, though intrinsically asymmetric, the dimer considered in the left panel of figure \ref{F:Qmap} pertains to the symmetric configurations of the trimer and hence there is no surprise that in some instances a dimeric level corresponds to a solitary level of the trimer rather than to an even-odd doublet.
If proved to occur in general,  the fact that such solitary levels are always even, as in figure \ref{F:Qmap}, remains to be understood.  

Notice that the (even) absolute maximum of the trimer for $\nu_{\rm t}<0$ seems to have a dimeric counterpart both in the left and in the right panel of figure \ref{F:Qmap}. Actually, as it becomes evident considering a larger value of $\tau_{\rm t}$, the overlap in the right panel is accidental, and hence it is not an exception to our discussion above.  This means that, as we mentioned in section \ref{S:MFmap}, one of the dimeric levels in the right panel  (the absolute maximum for $\nu_{\rm t}<0$) has no correspondent among the levels of the trimer. 

Owing to the quantum persistence of the dimer-trimer correspondence, most of the results about the asymmetric trimer presented in section \ref{S:dim} can be extended in a straightforward manner to the case of the trimer. 

As we mentioned at the beginning of the present section, figure \ref{F:Qmap} provides a quite useful key for interpreting the otherwise quite obscure trimeric version of figure \ref{F:dim}. Actually, owing to the quantum persistence of the dimer-trimer correspondence, most of the results about the asymmetric trimer presented in section \ref{S:dim} are so straightforwardly extended to the case of the trimer, that there is no need of including such figure in the present work. 

\section{Conclusions and perspectives}

We focused on the Bose-Hubbard Hamiltonians describing two small arrays of interacting BECs, the asymmetric dimer and the trimer.  
In section \ref{S:dim} we related some interesting features of the energy spectrum of the former Hamiltonian, whose asymmetry ensues from an imbalance between the on-site energy terms, to the energies of the fixed point of the corresponding mean-field effective Hamiltonian. These results are inspired by and generalize those concerning the symmetric dimer reported in reference \cite{A:Aubry}. Section \ref{S:trm}  is devoted to a similar investigation on the spectrum of the trimer. In order to find our bearings in the rather complex level pattern, in section \ref{S:MFmap} we exploited an interesting mapping among the mean-field dynamics of the trimer and that of the asymmetric dimer discussed in the previous section. In particular we showed that such mapping,  which is exact only for the integrable sub regime of the trimer, applies approximately also to the energies of most of its fixed points. Based on these mean-field results, in section \ref{S:Qmap} we observed that a similar approximate mapping exists also at the quantum level. We emphasize that the (quasi) dimeric character of the spectrum of the trimer is by no means evident from the structure of the relevant Bose-Hubbard Hamiltonian, and it strongly relies on the observations reported in  section \ref{S:MFmap}. The results gathered in sections \ref{S:MFmap} and \ref{S:Qmap}, along with the underlying systematic study of references \cite{A:PFLet1} and \cite{A:lungo}, allow to extend straightforwardly the observations reported in section \ref{S:dim}  to the case of the trimer.

The results presented in this work form the basis for a systematic study of the purely quantum trimer. Among the issues worth developing we mention a possible characterization of the quantum counterpart of chaos. Recall indeed that, owing to the non-integrability of the mean-field dynamics, the trajectories based in the vicinity of unstable fixed points are likely to display strong instabilities. Hence the mapping between fixed points and energy spectrum may provide a guide in recognizing possible  quantum manifestations of chaos.

Furthermore, if the system under concern is about to be realized experimentally, as promised by the recent advances in trapping techniques, the systematic study here proposed could provide a benchmark for verifying the effectiveness of the model Hamiltonians used to describe its physics. In this respect we recall that small bosonic lattices have been proposed for the implementation of quantum computation \cite{A:IoZa,A:PleBu}.

\section*{References}

\end{document}